# Research Directions in Network Service Chaining

Wolfgang John*, Kostas Pentikousis^, George Agapiou%, Eduardo Jacob¤, Mario Kind$,
Antonio Manzalini°, Fulvio Risso§, Dimitri Staessens&, Rebecca Steinert†, and Catalin Meirosu*

\* Ericsson Research    ^ EICT    % OTE Research    ¤ University of the Basque Country
$ Deutsche Telekom AG    ° Telecom Italia    § Politecnico di Torino    & iMinds    † SICS

Corresponding author email: *wolfgang.john@ericsson.com*

*Abstract*—Network Service Chaining (NSC) is a service deployment concept that promises increased flexibility and cost efficiency for future carrier networks. NSC has received considerable attention in the standardization and research communities lately. However, NSC is largely undefined in the peer-reviewed literature. In fact, a literature review reveals that the role of NSC enabling technologies is up for discussion, and so are the key research challenges lying ahead. This paper addresses these topics by motivating our research interest towards advanced dynamic NSC and detailing the main aspects to be considered in the context of carrier-grade telecommunication networks. We present design considerations and system requirements alongside use cases that illustrate the advantages of adopting NSC. We detail prominent research challenges during the typical lifecycle of a network service chain in an operational telecommunications network, including service chain description, programming, deployment, and debugging, and summarize our security considerations. We conclude this paper with an outlook on future work in this area.

## I. Introduction

Infrastructure network operators are currently struggling to meet growing user and traffic demands on their traditional connectivity services, for example, in terms of providing sufficient capacity and mobility support. While subscribers enjoy a constantly (and often drastically) declining "cost per bit", operator investments (CAPEX) and operational costs (OPEX) for the increasingly complex network infrastructure are rising: new technologies have to be incorporated while older investments are still operational and will be so for the foreseeable future. From a technical point of view, "over-the-top" service providers (OTT) can innovate and introduce new technologies at a rapid pace, while vendors close to the physical network enhance access technologies by orders of magnitude within a decade. This is only possible because the middle part of the protocol stack remains largely unchanged.

The price to pay for this approach, which calls for flexibility and innovation concentrating at the edges of the protocol stack, is the so-called "network ossification". Architectural kludges implemented through the introduction of middleboxes exacerbate this further: service chains must be carefully crafted from statically-assembled components chosen at design time. Then, once a network service is defined, little can change: operators can mainly perform minor configuration changes (e.g. parameter tuning) and address scalability through further infrastructure investment to reach more subscribers. In short, a whole network is purpose-built and optimized for a few static services. This modus operandi is advantageous in terms of service quality guarantees and has served, up to now, the telecommunications industry well. But it is particularly inflexible in the current market and technological conditions. Earlier investments in specialized hardware are difficult to re-tool and re-deploy with new functionality: network service providers (NSPs) cannot weave together best-of-breed technologies to form novel full service chains at will. In the end, NSPs operate, manage, and maintain costly and monolithic service silos deployed for decades.

To some extent, the current network service chaining (NSC) model is reminiscent of how mainframes were built in the early years of high-performance computing. For example, deployment models for advanced services, such as intrusion detection and prevention systems (IDS/IPS), firewalls, content filters and optimization mechanisms, deep packet inspection (DPI), caching, etc., are typically centered on monolithic platforms installed at fixed locations in or at the edge of the carrier core network. Besides being rigid and static, deployment of advanced network services and connectivity between network and service platforms often lack automatic configuration and customization capabilities, leading to significantly stretched deployment times and large operational complexity. Operational complexity is further aggravated by NSP organizational "silos" - separate teams and software systems manage particular network domains, treating service fulfillment and assurance as separate processes. As a result, troubleshoot times may vary greatly (from hours to days or even weeks).

Efforts to overhaul NSC gained significant traction recently in both research and standardization fora as NSPs seek to offer advanced services beyond basic connectivity, while optimizing infrastructure use and operational efficiency. For example, the IETF contemplates the creation of a dedicated working group on NSC. In this case, Quinn et al. [1] define a service chain loosely as *"the required functions and associated order that must be applied to packets and/or frames."* Conversely, Zhang et al. [2] motivate the need for *"steer[ing] traffic at the granularity of subscriber and traffic types"* and *"through the right inline service path"* but do not actually define NSC formally. In sum, earlier literature sketches the overall NSC context but falls short of providing a clear and concise definition of what NSC entails in detail.

We fill this gap in the following two sections and discuss our considerations regarding NSC in carrier-grade infrastructure networks. The remainder of this paper is dedicated to the current research directions in NSC, starting with service lifecycle aspects, including service chain description, programming, deployment, and debugging. We then introduce a holistic approach for tackling these issues in line with work planned in the EU-funded FP7 UNIFY project and conclude this paper with an outlook of the NSC area.

## II. MOTIVATION

Several drivers are expected to impact network and service infrastructure evolution in the coming years: technological progress in commercial of the shelf (COTS) hardware, cost reductions in processing and storage systems, growing availability of open source software defined networking (SDN) solutions, and "intelligence" migration towards user devices. These drivers will open new business opportunities and increase the competition in the ICT arena. The ossification of IP over transport networks hinders the flexible deployment of new network layer functionality (e.g. routing algorithms) or in-network security services (e.g. firewalls).

Changes to existing network platforms need careful engineering and customizations to address inter-dependencies between functional components and to meet high expectations on quality. Consequently, introducing new functionality into a deployed network is complicated, time-consuming and thus expensive. This rigidity provides few, if any, opportunities for re-tooling the network, and inhibits, in practice, the emergence of new revenue sources. In this context, reducing the "time-to-market" by minimizing the duration of the current network operator innovation cycle is critical.

Complexity compounds as third-party operators, such as wholesale customers, cellular, and cable network operators, call for access to the network platform and demand services beyond what the operator offers in retail. By the same token, access/aggregation carriers are no longer the only customer-facing infrastructure substrate and seek access to other access/aggregation operator's platforms in return, e.g. metro or cable network providers.

Significantly higher degrees of automation within management and configuration tasks could reduce both OPEX and CAPEX. OPEX reduction can be achieved, for example, by reducing the network touch-points (and thus possible configuration mistakes) and by assisting human administrators in configuring and managing equipment. CAPEX reduction can be achieved, for instance, by delaying network resource investments (e.g. by re-factoring and optimizing the use of available resources), through the virtualization of certain network functions that can run on standard data center hardware, and that can be instantiated in various locations, as considered by the ETSI Network Function Virtualization (NFV) [3] effort.

In today's network architectures, increasingly more complex services, such as IPTV, security services, and delivery optimizations, have been introduced through the deployment of middleboxes, both in the operator-controlled network as well as beyond the reach of the carrier in the home network environment. One example of an increasingly complex network platform is the 3GPP Evolved Packet Core (EPC) and its optimizations for content delivery and security. In addition to the typical eNodeB, S/PGW, MME and other network functions [4], an EPC deployment requires the following functions typically installed in independent boxes: i) Network Address Translation (NAT) from private IPv4 addresses to public IPv4/v6 addresses; ii) service access policing, e.g., for VPN, video platforms and VoIP; iii) infrastructure firewall protection; iv) a content distribution network (CDN) solution for efficient popular content distribution; and v) transcoding engines for optimized picture and video delivery.

In this case, customer traffic crosses several middleboxes, which effectively requires operators to define statically-provisioned service chains. Each middlebox is a stateful system supporting a very narrow set of specialized network functions and is based on purpose-built and typically closed hardware. This agglomeration of all kinds of middleboxes contributes to network ossification, and is responsible for a significant part of the network CAPEX/OPEX. As highlighted above, today an operator cannot re-use any of these middleboxes as they are carefully crafted to provide a single service: take one box out and the whole chain breaks. In contrast, the introduction of SDN and NFV in operator networks enables flexible allocation, orchestration and management of L2-L7 network functions and services and provides the substrate for dynamic network service chains.

Early-stage related work in SDN-based service chains originates in OpenFlow demonstrations. For example, OpenPipes [5] explored the possibility of using a modular network component system design in which self-contained in-network functions, such as digital image filters for video content distribution, could be called upon to create a video processing system. However, this work is basically a feasibility study and does not consider a carrier-grade network. Bari et al. [6] survey how virtualization can improve flexibility, scalability, and resource efficiency for data center operators and point to future research directions in that area, but provide few clues on virtualization benefits for telecommunication operators. On the other hand, work such as MobileFlow [7] details a possible way forward for introducing carrier-grade virtualization in EPC, but does not delve into NSC to a significant extent.

## III.   NETWORK SERVICE CHAINING

In general terms, we define dynamic Network Service Chaining as *a carrier-grade process for continuous delivery of services based on network function associations.* In this context, *continuous delivery* means dynamic network function orchestration and automated (re-)deployment used to improve operational efficiency. *Carrier-grade* means that the entire process is designed for high availability and fast failure recovery, with reliable testing capabilities integrated in every step of the process.

Fig. 1 illustrates how data travels from source to destination with and without the introduction of a dynamic NSC architecture. Today, each and every data packet has to be processed by a predefined series of (often hardware-based) "services" such as a security gateway service, a Deep Packet Inspection service (DPI), a firewall (FW), a load balancer, and so on. Fig. 1 is representative of what NSC principles can accomplish, resulting in a dynamic, software-configurable and upgradable system. NSC provides the means so that data can flow naturally without the intervention imposed by different services residing at different nodes. Network services can be implemented as part of a dynamic chain where each flow is processed by various service functions thus avoiding the need for deploying different physical network elements. Hence, NSC benefits from virtualization, allowing physical entities, whether manually configured or implemented with software, to be integrated seamlessly at a higher layers.

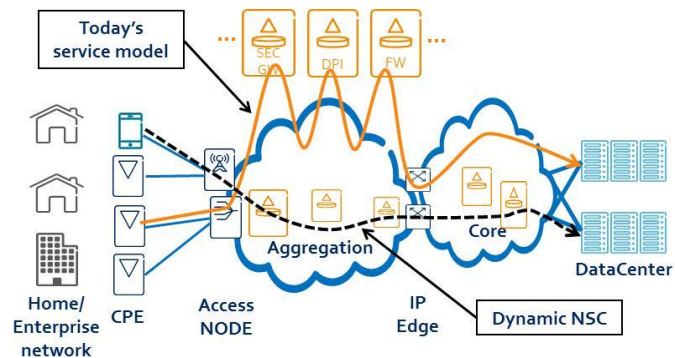

**Figure 1:** The bold orange line depicts traditional service creation models, following a predefined order of monolithic service elements. The dashed black line depicts dynamic NSC, passing physical and virtual service functions embedded into different network domains.

Another example of the benefits of adopting dynamic NSC is depicted in Fig. 2. When data travels in the same network (e.g. communication between two users who belong in the same network) it goes through different policy enforcement points, balancers, security gateways, traffic schedulers, etc., which may be in software or in hardware depending on the size and the complexity of the network. The goal of these elements is to provide better security and fairness to the end users. However, when traffic has to traverse different network domains, additional operator investment in terms of software and hardware are required, e.g. in the form of a provider edge (PE) gateway which consists of software (different policy elements) and hardware for routing and forwarding the traffic. In these cases, things become more complex and data may experience heavy deterioration in terms of delay depending on the cross-domain network load and the number of the policy elements through which this data is passing. With dynamic NSC implemented in each of the network domains and by exploiting the NSC functionality on the PE elements, it will lead to more intelligent traffic steering and thus provide traffic performance acceleration. Sensitive data and multimedia flows can cross different networks in a reliable and, more importantly, predictable manner when NSC is implemented in the edge of the different network domains.

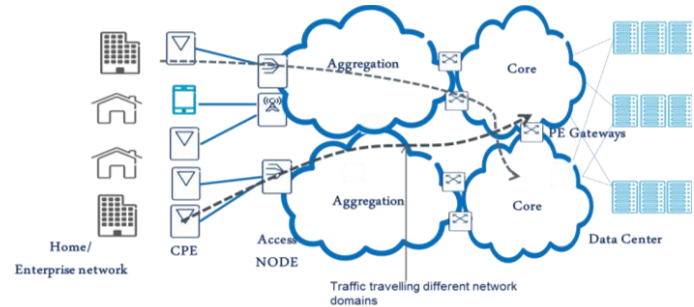

**Figure 2** Service provider network interconnection

However, dynamic service chaining and the evolution of network virtualization from data centers into carrier networks do not come without their own challenges. Due to the dynamic nature of the service path, it will no longer be feasible to allow for lengthy discovery processes separating the fulfillment and assurance steps. In addition, each network service chain could evolve automatically to include new service components, if necessary, or shed components that are no longer needed at run time. The flexible bundling of service components customized to individual subscribers will lead NSC operators to manage large numbers of services and service instances. This is unlike the current operational environment, where carriers manage dozens of services, which apply to millions of subscribers. To be able to handle large numbers of flexibly created dynamic network services, we define *continuous network service delivery* as the operator ability to introduce customized services at a rapid pace while maintaining carrier-grade end-user quality of experience.

## IV. NSC RESEARCH DIRECTIONS

In order to enable dynamic NSC in future operator networks, several challenges need to be addressed. This section follows the lifecycle of a compound service realized via NSC, which includes description and programming, service instance deployment, continuous network service delivery, as well as security, and identifies the associated research directions.

### A. NSC Description and Programming

Network service chains can be considered as particular cases of service composition. As a research topic, service composition has been studied extensively before [8][9]. Research questions which are still open include optimization strategies for decomposition and aggregation of services and service blocks, service modeling languages, design for personalization, mobility, context awareness and adaptation, modeling and enforcement of policies, risk and trust.

Although every link in the service chain could be treated as a service on its own, it is yet unclear to what extent the *aggregation of service blocks* needs to be performed through interfaces that are highly descriptive or whether simple REST-ful interfaces might be more appropriate for the task. There are similarities between characteristics that are desirable for chain links and the netlet concept from the NENA architecture [10].

Recent work in the domain of network programming languages (such as Pyretic [11] and Maple [12]) shows how to implement network functionality by controlling the flow space in an SDN/OpenFlow switch in a programmatic manner. Netcore [13] allows policies to be described in terms of arbitrary functions that cannot be directly realized on physical switches. To handle such policies, the compiler generates an underestimation of the overall policy using a simple static analysis, and then uses partial evaluation to refine this underestimation at run time using the actual packets seen in the network. Less generic solutions, such as SIMPLE [14], have been proposed to address challenges related to mapping towards physical resources and controller visibility into the functionality exposed by a middlebox.

The environment of a dynamic *service chain* calls for *programming languages* that address complex policies in accordance to the packet processing capabilities in the chain links. For instance, complex functionality, such as caching or intrusion detection, needs programmability constructs that go beyond simple manipulation of flow tables and address handling. As service chains will be deployed on both physical and virtual infrastructures, it is reasonable to believe that virtual switches may in time develop characteristics that are beyond the reach of their physical counterparts in terms of flexibility, feature complexity and frequency in release cycles.

Ways to accurately describe the service characteristics are needed in order to enable automated NSC deployment and optimization. *Service description* needs to cover both the service level and resources involved, from hardware (or a virtual representation) and software point of view. The problem of accurately describing high-level services has been tackled from different angles. The Grid community has focused its efforts on Semantic Grid and OGSA [15]. The cloud computing community has been working on the topic too [16]. Regarding resource description languages with a network orientation, some work is available in the literature (e.g. VXDL[17]). From a NSC perspective, VXDL includes temporal constraints difficult to synchronize between orchestration engines, and not directly supported by resources.

Dynamic service chains are expected to be instantiated in large numbers, likely in the order of the number of subscribers to a particular network. Service and resource description languages need to address such *scalability aspects* natively in order to allow for efficient deployment. Beyond simple temporal constraints, other constraints related to QoS, resource sharing and mobility, security and energy efficiency need to be supported. Particular attention needs to be paid to describing network flows (in OpenFlow terms) that must be forwarded between elements. This task is not trivial because the OpenFlow flow-match-action definition is so rich (and starting with version 1.2, expandable) that the possibilities for aggregating, dividing and defining flows are almost endless.

Virtualization technologies enable resource sharing in a transparent manner between multiple service chain instances. In an OpenFlow context, the OFELIA Control Framework [18], Layer 2 Prefix-based Network Virtualization [19], FlowVisor [20] and FlowN [21] are examples in this respect. Dynamic flow reconfiguration at the architectural level of the infrastructure by different actors owning different service chain instances will have a large impact on scalability requirements (mainly in terms of signaling) which must be addressed. Virtual machine migration can, too, impact behavior at the flow level and lead to sub-optimal resource utilization when the virtual and physical infrastructure descriptions are not able to encompass all applicable constraints. Maintaining optimal resource usage under such dynamic conditions requires adaptive monitoring and optimization approaches, crucial for tracking and optimizing resources between multiple service instances relative to usage and policy constraints.

The definition of a dynamic service chain needs to facilitate *monitoring and problem troubleshooting* for chain instances under live operations. Certain steps were made in this direction, ranging from more theoretical proposals such as [22], to specifying a set of key performance indicators as part of the Service Measurement Index [23] and developing Application Programming Interfaces (API) such as the TMForum Simple Management API [24]. The exchange of service monitoring information is generally well understood and challenges relate more to the shear amount of information to be provided as well as privacy concerns. However, as illustrated by the discussion in the IETF ALTO working group regarding privacy requirements in the exchange of topology data [25], certain information that could be used to facilitate troubleshooting is considered sensitive by the providers. More than a simple modeling exercise, determining what troubleshooting-related information needs to be made available between links in a dynamic service chain requires establishing a balance between the utility of every bit exposed versus the potential business risks. However, in order to facilitate troubleshooting via automated tools, fairly detailed information as well as troubleshooting-related actuators need to become available through programmatic interfaces.

*B. Service Instance Deployment*

Datacenter networking is typically based on rather homogeneous platforms. In stark contrast, telecommunications gear has traditionally used a mixture of different components, such as network processors, ASICs, and a wide variety of processing elements. Heterogeneity limits platform openness to general programmers. In practice, it is difficult to allow anyone but the network equipment manufacturer to create, install and deploy software on physical devices.

Future network equipment, suitable for NSC, could follow the SDN datacenter model, i.e. migrate toward a *universal node* paradigm, in which the computation and storage resource architecture is mutated from the standard high-volume hardware deployed in datacenters. This would lead the entire path from network edge to datacenter to be seen as a homogeneous programmable platform, enabling software deployment at any place of this (long) programmable path. Similarly, currently monolithic (and complex) functions can be split into several components, each one running at the location that is the best suited for the overall service operation.

As an illustrative example, consider a complex function running on current networks such as a Broadband Remote Access Server (BRAS), which is typically implemented as a dedicated network element with deeply integrated monolithic software. NSC based on universal nodes, on the other hand, would allow function refactoring into modules, with some of them executed at the network edge (e.g., user session termination), others in the core (e.g., content caching), and the rest in the datacenter (e.g., user authentication). From the network operator point of view, this NSC-based BRAS is a unique function, without any reference to the exact physical location of different modules. In principle, automatic dispatchers would optimize the location of each component of this function by (dynamically) relocating each module based on different parameters such as its CPU/memory requirements, the network traffic generated by the communication between different modules, and so on.

While service chaining distributed across the operator network and datacenters is a likely way forward, future physical architectures of network devices are not clear yet, which opens up an entire area of research. For instance, one option is to have virtual network devices that result from the aggregation of several distinct components, such as a network switch combined with a set of traditional servers with processing and storage resources. A second option would be to integrate diverse resources (e.g., components specifically targeting network tasks such as network processors or ASICs, plus general purpose hardware such as mainstream CPUs, memory, etc.) in each device. A third option would avoid altogether network-specific components in deployed gear, assuming that the overall performance through the use of solely general purpose hardware is acceptable. In fact, it is worth mentioning that network functions may also include data plane components, i.e., modules that need to inspect, and potentially modify, large amounts of network traffic and therefore benefit from dedicated hardware accelerators in network devices.

A second important research question is independent from the physical architecture of future network devices and relates to the *modular design of network functions*. For instance, at one end of the spectrum we may have a single, monolithic function that can be installed at any network location. On the other end, we can imagine a highly granular partition of the same function into very small components such as regular expression matching, lookup table processing, etc., with each one operating at a different location of a programmable path.

Finally, *mapping service chain components to available resources* in the network is still an open research topic. A mapping function needs to determine where certain blocks may be installed. This component may require additional performance monitoring models that measure/predict resource state and provide input to the mapping functions. When multiple mappings are possible, an optimal solution imposing constraints on the amount of resources to be reserved for these blocks (CPU cycles, memory, network interfaces, physical location and so on) could be chosen, or some other policy may be selected. Such optimization problems have long been known as NP-hard ([26][27][28]) and various heuristics were developed to make them computationally tractable. Methods based on probabilistic approaches, capable of accounting for uncertainty and variations in the network, are promising in this respect [29][30][31].

*C. Continuous Network Service Delivery*

Service chains may be assembled manually through a user interface or dynamically via algorithmic development. Either way, operators can no longer afford extensive field trials before introducing new services and changes. Instead, they need to be empowered with a toolbox that facilitates daily operations and troubleshooting, in a fashion similar to the DevOps tools gaining popularity in the IT world [32]. DevOps includes tools common to both development and operations teams. In SDN, the connection between implementing network policies and easily determining the source of performance problems was highlighted by Kim and Feamster [33].

DevOps borrows from agile software development methodologies in order to facilitate cross-team communication and greatly increase infrastructure automation. In this context, *workflow definition* for testing, validation and troubleshooting may be considered a challenge. As discussed in [34], even software-defined networks require a fairly complex and time-consuming constructed workflow in order to troubleshoot network functionality using state of the art tools. Dynamic service chains need such workflows to be tailored to their needs which can evolve dynamically. Therefore we need mechanisms to track and incorporate such changes in troubleshooting processes and tools.

NSC also calls for modern *model checking methods* as testing must be performed before a service instance is activated and delivered to the customer. Due to the dynamic properties of the chain, currently defined static methods used today for checking, e.g. connectivity services (ITU T-Y.1564), are not suitable. Software development model checking techniques have recently been employed in an SDN context [35]. However, model checking for dynamic service chains ought to address a series of challenges in terms of the number of instances that could be tested simultaneously on a given infrastructure and duration of tests. The effectiveness of such approaches is currently limited by the state-space explosion due to the composition of service chain segments. Moreover, it would be natural that such checkers are generated on demand and automatically configured to account for policy restrictions for particular chains. This line of work appears very promising as we move forward in NSC research.

As service chains are deployed, we need an infrastructure that dynamically keeps track and can "zoom in" on particular components that could be problematic within a chain, i.e. we would like to have *programmable observation points* within each chain covering both network and service components. With programmable, we mean that the observation points are embedded within the service chain at service definition time alongside other policies. As a result, the infrastructure can perform situation analysis in an autonomic fashion, determine the exact context, decide and react to changing conditions, e.g. when to implement the service instance migration.

Furthermore, programmability also means the possibility to programmatically *assemble basic observation constructs* (such as counters, active measurement capabilities, etc.) into tailor-made tools that can detect specific problems at various service chain components. This type of functionality obviously requires tradeoffs between security, confidentiality, visibility, and resource consumption when it comes to infrastructure internal observation capabilities and the level of exposure toward the customer on the service chain interface. Furthermore, the required dynamicity should be implemented with low signaling overhead towards the infrastructure. Information will be mapped to the troubleshooting workflow, summarized and presented in a manner that makes it easier for human operators to take a decision on further actions.

With respect to the vantage point over the entire set of service chains deployed in the field, *scalable observability* is a major concern. Wide-scale deployment of programmable observation points may potentially lead to huge amounts of monitoring data. In practice, for highly dynamic large-scale systems, a reasonable tradeoff is to avoid reflecting the exact network state information continuously and at a very fine granularity. Network state can be approximated, reaching a balance between estimation accuracy and degree of tolerable uncertainty. In other words, capturing the network state in an accurate, efficient and scalable manner requires monitoring components that are adaptive to changing network conditions - a promising research area in NSC. Scalable and flexible tools for SDN fault management and performance monitoring also require efficient real-time data processing to handle massive amounts of network data, for example by combining probabilistic approaches [36] and big data analytics.

*D. Security Considerations*

NSC introduces interesting security research topics for different reasons. From a classic security domain point of view, a service chaining process (which takes place in the provider domain) might be triggered by an end user (which is in another domain). Note that today the user domain relies on various legacy access networks that connect to the core, where connectivity is granted after an authentication process such as PPP or serial number matching, with further services requiring additional authentication/authorization processes. These techniques are not very well suited to services composed through dynamic NSC as discussed in this paper.

Short-lived network services, for instance upgrading connectivity performance for a given period (e.g. to offer a "premium" service in terms of network delay for supporting online games while one is connecting to a game central), bringing the service to targeted users in a single subscriber network (e.g., to benefit only one of the home users), or supporting user-service mobility (watching on-demand movies using one's own subscription while visiting friends at their home), are difficult to support with the current schemes. Security demands increase if we take into account the deployment of these services not only on the subscriber's access network, but on visited access networks too. Additional services that will rely on a richer definition of security services include delegating rights to other users, e.g., granting access to a service bundle to another user while the subscriber is travelling. A security architecture for dynamic NSC will have to deal not only with legacy and lower layer protocols used in access networks today, but also shift from a (CPE) device-based authentication model to a user identity based one. To support this vision, research around mixed private/public key architectures that will be able to cope with the huge number of users and interactions expected in dynamic NSC will need to be undertaken.

In the provider domain, the use of static services with proprietary equipment has been beneficial in the past in terms of security. The dynamic aspect of NSC implies that new effort will be needed in deployment design. In fact, it is widely recognized that virtualization technologies can help to control the scope in which services are deployed. Finally, as Quinn et al [1] already detect, but do not elaborate on, there are clear security implications on data and control plane management when deploying NSC which future research ought to consider.

V. FUTURE WORK

After detailing the relevant research directions in NSC, we take the opportunity to introduce the EU-funded FP7 project UNIFY, which sets out to tackle many of the issues mentioned above through a holistic approach [37]. The core project aim is to provide the means for flexible service creation within the context of unified cloud and carrier networks, especially focusing on network functions. Specifically, UNIFY finds current carrier networks to be slow, rigid in terms of functions and resources, and inflexible with respect to service creation. Thus, UNIFY envisions an architecture where the entire network from home devices to data centers forms a unified production environment. As a consequence, an NSP can distribute functions and state anywhere in the network, aided by automated orchestration engines. In other words, UNIFY envisions an automated, dynamic service creation platform, leveraging fine-granular NSC.

To accomplish such a unified production environment, the project will focus on four key aspects. First, UNIFY will consider the network services for a converged fixed-mobile network and study the decomposition of these traditional network functions into more fine granular components. UNIFY will identify the minimal set of components which, once in place, can provide more flexibility for network service chaining. Second, UNIFY will define a service abstraction model and a proper service creation language suitable for dynamic NSCs. This includes aspects dealing with orchestration and network function placement optimization through novel algorithms, enabling the automatic placement of networking, computing and storage components across the infrastructure. Third, in the framework of Service Provider DevOps, new management technologies will be developed, based on the experience from data centers, and integrated into the orchestration architecture, addressing the challenges of dynamic service chaining. Finally, UNIFY will evaluate the applicability of a universal node based on commodity hardware in order to support both network functions and traditional data center workloads, with an investigation of the need of hardware acceleration.

## VI. Conclusion

This paper discussed network service chaining in the context of future infrastructure networks. After illustrating how service chains are crafted today, we motivated the need for dynamic NSC and presented how service chains can be employed in future operator networks in order to provide cost reductions, increased flexibility, and time-to-market acceleration throughout the network. We then went through our design considerations for NSC, including the key role it can play in accelerating the design, implementation and deployment of novel service offerings in infrastructure networks as well as the potential for carrier CAPEX and OPEX reduction. The key contribution of this paper is a detailed array of research directions in the context of NSC, including service instance deployment, network service definition, programming, and operations, as well as the concept of continuous network service delivery. Finally, we summarized how the EU- FP7 UNIFY project aims to address some of the research challenges introduced in this paper.

Dynamic NSC is a very promising research area with several topics that will need to address challenges which hitherto were unknown in telecommunications networks. For example, service chain "debugging" in an NSC era and the corresponding network fault isolation processes today entail completely different aspects. Therefore, research results towards dynamic NSC will have significant impact in the way we design, operate, and maintain networks in the coming years. Furthermore, in contrast to major players in datacenter networking, network infrastructure carriers prefer solutions that are interoperable and have global reach and applicability. As such, we also expect that as research in NSC matures and is demonstrated to work well in practice, some of the NSC focus will be diverted towards interoperable solutions across operator networks and thus international standardization.

## Acknowledgments

This work was conducted within the framework of the FP7 UNIFY project, which is partially funded by the Commission of the European Union. Study sponsors had no role in writing this report. The views expressed do not necessarily represent the views of the authors' employers, the UNIFY project, or the Commission of the European Union. We thank Fritz-Joachim Westphal, Ioanna Papafili, Róbert Szabó and Pontus Sköldström for constructive comments as well as all partners who contributed in the UNIFY preparation phase.